\documentclass[twocolumn,showpacs,preprintnumbers,amsmath,amssymb]{revtex4}

\usepackage{graphicx}
\usepackage{dcolumn}
\usepackage{bm}

\begin{document}
\title{\bf{The states of W-class as  shared resources for \\ perfect teleportation and superdense
coding}}
\author{Lvzhou Li, Daowen Qiu}%
\email{issqdw@mail.sysu.edu.cn (D. Qiu).}
 \affiliation{%
 Department of
Computer Science, Zhongshan University, Guangzhou 510275,
 People's Republic of China
}%

\begin{abstract}
 As we know, the states of triqubit systems have two important classes: GHZ-class and W-class.
 In this paper,  the states of W-class are considered for
 teleportation and superdense coding, and are generalized to
 multi-particle systems. First we  describe two transformations of the shared resources for teleportation and superdense
 coding, which allow many new protocols from some known ones for that. As an application of these transformations, we obtain
 a
sufficient and necessary condition for a state of W-class being
suitable for perfect teleportation and superdense coding. As
another application, we find that state
  $|W\rangle_{123}=\frac{1}{2}(|100\rangle_{123}+|010\rangle_{123}+\sqrt{2}|001\rangle_{123})$
  can be used  to transmit three
  classical bits by sending two qubits, which was considered to be impossible by P. Agrawal and A. Pati [Phys. Rev. A to be published].
  We generalize the states of
  W-class to multi-qubit systems and multi-particle systems with  higher
  dimension. We propose two protocols for teleportation
  and superdense coding by using W-states of multi-qubit systems that generalize the protocols by using
  $|W\rangle_{123}$ proposed by P. Agrawal and A. Pati. We obtain an optimal way to partition   some W-states
  of multi-qubit systems  into two subsystems, such that the entanglement between them achieves
  maximum value.
  \end{abstract}
\pacs{03.67.-a, 03.65.Bz} \maketitle
  \section{\label{Introduction}Introduction}
 D\"{u}r {\it et al} \cite{Dur00} pointed out
that the states of three qubits can be entangled in two
inequivalent ways: GHZ-class and W-class.  States in one class can
not be converted from states  in the other class  by stochastic
local operations and classical communication (SLOCC) \cite{Ben99}.
With respect to loss of qubits, the two classes are rather
different. The states of W-class are robust against loss of qubits
(i.e., if we trace out any one qubit, then there is some genuine
entanglement between the remaining two qubits), while GHZ-states
are not. The GHZ-class has been extensively  studied from many
aspects, while W-class may need more effort to clearly
characterize it.

Recently, more attentions have been taken on the states of
W-class. They have been considered for many important quantum
information processing tasks \cite{Lee02,Gor03,Joo03,Pan06,Cab02}.
W-states were considered as quantum channel for teleportation of
entangled pairs in \cite{Gor03}. Probabilistic  teleportation of a
qubit state via a W-state was  studied in \cite{Joo03}.
Furthermore, in \cite{Pan06} the authors discovered a subclass of
W-class suitable for perfect teleportation and superdense coding,
but they did not give the sufficient and necessary condition for a
W-state being suitable for that. In addition, W-class has been
used for quantum key distribution \cite{Lee02}, and  in
illustrating violation of local realism \cite{Cab02}. At the same
time, there have been various proposals to prepare W-states
\cite{Guo02}. Considering the importance of W-class, in this paper
we focus on the states of W-class for perfect teleportation and
superdense coding.

As we know, quantum teleportation and superdense coding are two
amazing and interesting processes in quantum information theory,
where entangled states as shared resources play a crucial role
\cite{Ben92,Ben93}. The two processes have a close relationship
that has been investigated by Werner \cite{Wer00}. In the original
protocols \cite{Ben92,Ben93} for teleportation and superdense
coding, EPR pairs were considered as the shared resources. Latter,
more candidates have been considered. For instance, maximally
entangled states of triparticle  were considered in
\cite{Cer01,Mar00, Gor00}, and maximally entangled states of
multi-particle were considered in \cite{Bos97, Long, Bru05}.
Specially, non-maximally entangled states have been considered for
teleportation and superdense coding by many authors. For instance,
non-maximally entangled states have been considered for
probabilistic teleportation \cite{Li00,Agr02, Pat04, Gor06} and
probabilistic superdense coding \cite{Guo00, Pat05}. In addition,
the classical information capacity of deterministic superdense
coding  with non-maximally entangled states has been  studied by
\cite{Hau96,Bow01,Moz05}.  Recently, various kinds of quantum
channels have been explored for deterministic and unambiguous
superdense coding \cite{Wu06,Fen06,Fan06}.

 Observing the states of
triparticle, interestingly we find that some states from GHZ-class
are suitable for perfect teleportation and superdense coding, but
some states are not. For instance, there are two states of
GHZ-class:
\begin{align}
&|GHZ\rangle_{123}=\frac{1}{\sqrt{2}}(|000\rangle_{123}+|111\rangle_{123}),\\
&|\widetilde{GHZ}\rangle_{123}=\frac{1}{\sqrt{3}}(\sqrt{2}|000\rangle_{123}+|111\rangle_{123}).
\end{align}
It is well known that the first state can be used for perfect
teleportation and superdense coding.  But the second one can not.
At the same time, similar case is found in W-class. For the
following two states of W-class:
\begin{align}
&|W\rangle_{123}=\frac{1}{2}(|100\rangle_{123}+|010\rangle_{123}+\sqrt{2}|001\rangle_{123}),\label{3}\\
&|\widetilde{W}\rangle_{123}=\frac{1}{\sqrt{3}}(|100\rangle_{123}+|010\rangle_{123}+|001\rangle_{123})
,
\end{align}
the first state can be used for teleportation of a qubit state and
for superdense coding of  two classical bits by  one qubit as
shown in \cite{Pan06}, but the second state can not.

In this paper, we focus on the states of W-class. Firstly we have
a brief analysis on the above two states from W-class, to find out
what difference lies  between them.
 We can partition the three particles whose state is $|\widetilde{W}\rangle_{123}$  into two subsystems
 in three ways: $1|23$, $2|13$, and $12|3$.  Then we have
\begin{align}
\begin{split}
&\rho_1=\rho_2=\rho_3=\frac{2}{3}|0\rangle\langle0|+\frac{1}{3}|1\rangle\langle1|,\\
&\rho_{23}=\rho_{13}=\rho_{12}=\frac{2}{3}|\psi\rangle\langle\psi|+\frac{1}{3}|00\rangle\langle00|,
\end{split}
\end{align}
where $|\psi\rangle=\frac{1}{\sqrt{2}}(|01\rangle+|10\rangle)$ is
an EPR pair.
 If we calculate the entanglement   between the two subsystems
 resulted
from the above partitioning ways, we have
\begin{align}
\begin{split}
E_{1(23)}(|\widetilde{W}\rangle_{123})&=E_{2(13)}(|\widetilde{W}\rangle_{123})=E_{(12)3}(|\widetilde{W}\rangle_{123})<1.
\end{split}
\end{align}
Note that here we adopt the definition of partial entropy
entanglement, where the entanglement between subsystems A and B
involved in the pure state $|\Psi\rangle_{AB}$ is defined as
\begin{align}
E_{A|B}(|\Psi\rangle_{AB})=S(\rho_A)
\end{align}
where $\rho_A=tr_B(|\Psi\rangle\langle\Psi|)$ and $S(\rho_A)$ is
the von Neumann entropy \cite{Nie}.

 Now the above results  show that: (i) State $|\widetilde{W}\rangle_{123}$ has
symmetry property  that no matter how we partition it into two
subsystems, the results are always the same. (ii) The entanglement
between any two subsystems resulted from the above partition is
less than one ebit, and therefore it is not suitable for perfect
teleportation and superdense coding. (iii) If any qubit of the
state is lost, the residual two qubits still share genuine
entanglement between them, which is called the {\it robustness}
against loss of qubits. Contrary to that,  the states of GHZ-class
do not have this robustness.

For  state $|W\rangle_{123}$, we  can also partition it into two
subsystems by the same way stated before, and then we have
\begin{align}
\rho_3=\frac{1}{2}|0\rangle\langle0|+\frac{1}{2}|1\rangle\langle1|,
\hskip 1mm
\rho_{12}=\frac{1}{2}|\psi\rangle\langle\psi|+\frac{1}{2}|00\rangle\langle00|,
\end{align}
and
\begin{align}
\begin{split}
&\rho_1=\rho_2=\frac{3}{4}|0\rangle\langle0|+\frac{1}{4}|1\rangle\langle1|,\\
&\rho_{23}=\rho_{13}=\frac{3}{4}|\phi\rangle\langle\phi|+\frac{1}{4}|00\rangle\langle00|,
\end{split}
\end{align}
where
$|\phi\rangle=\frac{|10\rangle+\sqrt{2}|01\rangle}{\sqrt{3}}$ is a
partial entangled pair. Calculating the entanglement between any
two subsystems resulted, we have
\begin{align}
E_{(12)3}(|W\rangle_{123})=1,
E_{1(23)}(|W\rangle_{123})=E_{2(13)}(|W\rangle_{123})<1.
\end{align}

 From the above results we know that: (i) If we partition the state in the
 way: $12|3$, then the two subsystems resulted share an  entangled
 state with one ebit of entanglement. Therefore, if  Alice has
 particle `3' and Bob has particles `1' and `2', then Alice can send
 Bob two classical bits of  information  by sending one qubit, or Bob  can
  teleport a state of qubit to Alice as shown in \cite{Pan06}. (ii) The state does  not have  symmetry property, and thus the other two partitioning ways
 lead to two subsystems whose entanglement is less than one ebit.
  (iii) The state also has the robustness against
 loss of qubits.

In summary, from the above considerations, we have seen that
except for the robustness against loss of qubits, the states of
W-class have also some other interesting properties. For instance,
state $|\widetilde{W}\rangle_{123}$ has symmetry property that may
be very useful for some applications, but the entanglement
involved does not attain one ebit, which, on the other hand, may
be its limitation. Oppositely, state $|W\rangle_{123}$, without
symmetry property, can be used as a maximally entangled state if
we choose an appropriate way to partition it. Therefore, the
states of W-class like $|W\rangle_{123}$ and
$|\widetilde{W}\rangle_{123}$ may be worthy of further
consideration, and we would like to do that in this paper.

In this paper, based on the previous works \cite{Dur00,Pan06}, we
characterize further the states of W-class, and consider further
the  protocols  for perfect teleportation and superdense coding
via W-class. In Section \ref{Protocols}, our main aim is to find
what states of W-class can be used for teleportation and
superdense coding. Firstly, in Sec.~\ref{Protocols:1} and
\ref{Protocols:2}, we describe two transformations of the shared
resources for teleportation and superdense coding, which will
allow many new protocols for that from some known protocols. In
Sec.~\ref{Protocols:3}, as an application, we obtain a sufficient
and necessary condition for a state of W-class being suitable for
perfect teleportation and superdense coding. In
Sec.~\ref{Protocols:4}, as another application, we find that state
$|W\rangle_{123}$ can be used to transmit three classical bits by
sending two qubits. In Section \ref{Characterization}, our aim is
to generalize states $|W\rangle_{123}$ and
$|\widetilde{W}\rangle_{123}$ to multi-particle systems, and
characterize them.
 In Sec.~\ref{Characterization:1}, we generalize states $|W\rangle_{123}$ and
$|\widetilde{W}\rangle_{123}$ to multi-qubit systems. In
Sec.~\ref{Characterization:2}, we propose two protocols for
teleportation and superdense coding that generalize the protocols
by using state $|W\rangle_{123}$ indicated in \cite{Pan06}. In
Sec.~\ref{Characterization:3}, for the states that generalize
$|\widetilde{W}\rangle_{123}$ to N-qubit systems, we point out an
optimal way to partition them into two subsystems such that the
two subsystems resulted share maximum entanglement, and, when $N$
is an even number, in a certain sense these generalized states can
be exploited  as maximally entangled states. In
Sec.~\ref{Characterization:4}, we generalize state
$|W\rangle_{123}$ to multi-particle systems with higher dimension.
Finally, some concluding remarks are made in Sec.~\ref{Con}.

 \section{\label{Protocols}Protocols for superdense coding and teleportation}
In this section, firstly we will describe two transformations of
the shared resources for superdense coding and teleportation. Then
as an application of these transformations,  we will obtain a
sufficient and necessary condition for a state of W-class being
suitable for perfect teleportation and superdense coding. Also, we
will find that state $|W\rangle_{123}$ can be used to transmit
three classical bits by sending two qubits, which was considered
to be impossible in \cite{Pan06}.

\subsection{\label{Protocols:1}Transformations of the shared resources for superdense coding}
The standard protocol for superdense coding can be described in
the following.  Suppose that Alice possesses  subsystem A and Bob
possesses subsystem B and the two subsystems share an entangled
state $|\varphi\rangle_{AB}$.  Alice can apply operators from the
set of unitary operators  $\{U_A^x\}$ on her subsystem A, and then
send A to Bob. Then the states of bisystem AB belonging to Bob
will form an orthogonal set $\{|\Phi_x\rangle\}$. That can be
described as
\begin{align}
(U_A^x\otimes I_B)|\varphi\rangle_{AB}=|\Phi_x\rangle.
\end{align}
Because of the orthogonality of  set $\{|\Phi_x\rangle\}$, Bob can
make a  projective measurement on system AB with projectors
$P_x=|\Phi_x\rangle\langle\Phi_x|$ to perfectly distinguish the
 set $\{|\Phi_x\rangle\}$, such that Bob can know exactly
which operator Alice has applied. In this process, if the number
of the operators in set $\{|\Phi_x\rangle\}$ is $N$, then Bob can
get $\log_2 N$ classical bits of information from Alice.

If we take a viewpoint from  discrimination between unitary
operations \cite{Ac01,Ari01}, then the core of superdense coding
is to find out as many unitary operators as possible, such that
they can be perfectly discriminated  by  state
$|\varphi\rangle_{AB}$.

Now suppose there is another   state $|\varphi^{'}\rangle_{AB}$
given by
\begin{align}
|\varphi^{'}\rangle_{AB}=(U_A\otimes
I_B)|\varphi\rangle_{AB}\label{Ts(i)}
\end{align}
where $U_A$ is a unitary operator acting on subsystem A and $I_B$
is identity operator on B. Then we have
\begin{align}
\begin{split}
(U_A^xU_A^{\dagger}\otimes
I_B)|\varphi^{'}\rangle_{AB}&=(U_A^xU_A^\dagger\otimes
I_B)(U_A\otimes I_B)|\varphi\rangle_{AB}\\
&=(U_A^x\otimes I_B)|\varphi\rangle_{AB} \\
&=|\Phi_x\rangle. \label{Tsr(i)}
\end{split}
\end{align}
Thus it is shown that by sharing state $|\varphi^{'}\rangle_{AB}$,
Alice and Bob can also fulfill the  same task as by sharing state
$|\varphi\rangle_{AB}$, and the only thing changed is that the set
of operators applied by Alice turns to $\{U_A^xU_A^\dagger \}$.

We can also take a unitary operation $V_B$ on subsystem B getting
a new entangled state in the following way:
\begin{align}
|\varphi^{''}\rangle_{AB}=(I_A\otimes V_B)|\varphi\rangle_{AB}
\label{Ts(ii)}.
\end{align}
Then we have
\begin{align}
\begin{split}
(U_A^x\otimes I_B)|\varphi^{''}\rangle_{AB}&=(U_A^x\otimes
I_B)(I_A\otimes V_B)|\varphi\rangle_{AB}\\
&=(I_A\otimes V_B)(U_A^x\otimes I_B)|\varphi\rangle_{AB}\\
&=(I_A\otimes V_B)|\Phi_x\rangle. \label{Tsr(ii)}
\end{split}
\end{align}
From the orthogonality of set $\{|\Phi_x\rangle\}$ and the
unitarity of $V_B$,  the states in set $\{(I_A\otimes
V_B)|\Phi_x\rangle\}$ are clearly mutually orthogonal. Therefore,
by sharing  state $|\varphi^{''}\rangle_{AB}$, Alice and Bob can
also fulfill the same task as by sharing $|\varphi\rangle_{AB}$.
The only change is that Bob should make a projective measurement
on AB with projectors $P_x=(I_A\otimes
V_B)|\Phi_x\rangle\langle\Phi_x|(I_A\otimes V_B^\dagger)$.

It is worth pointing out that transformations \eqref{Ts(i)} and
\eqref{Ts(ii)} can occur simultaneously. Then  we  should combine
\eqref{Tsr(i)} and \eqref{Tsr(ii)} getting a new protocol. For
simplicity, we describe them separately. Similar case  will occur
in the protocol for teleportation as  we will show in
Sec.~\ref{Protocols:2}.

\subsection{\label{Protocols:2}Transformations of the shared resources for teleportation}
The standard protocol for teleportation can be described as
follows. At first, Alice and Bob share an entangled state
$|\varphi\rangle_{AB}$ of bisystem AB, of which subsystem A
belongs to Alice and subsystem B belongs to Bob. Also Alice
possesses another system `a' whose state is $|\Psi\rangle$. Now
Alice's task is to send the state $|\Psi\rangle$ to Bob such that
the state of Bob's subsystem turns to that, with the help of the
shared state $|\varphi\rangle_{AB}$ and some classical
communication between them. This task can be fulfilled, if
 $|\Psi\rangle_a|\varphi\rangle_{AB}$ can be rewritten in the
 form
 \begin{align}
|\Psi\rangle_a|\varphi\rangle_{AB}=\frac{1}{\sqrt{D}}\sum_{x=1}^D|\Phi_x\rangle_{aA}U_x|\Psi\rangle_B
 \end{align}
 where $\{|\Phi_x\rangle_{aA}\}$ is a set of mutually orthogonal
 states of joint system `aA', and $\{U_x\}$ is a set of unitary
 operators on subsystem B. Alice can now fulfill her task by two
 steps:
 \begin{itemize}
    \item  Make a projective measurement on `aA' in
     a basis that includes $\{|\Phi_x\rangle_{aA}\}$,
     getting  measurement result `$x$' with probability
     $\frac{1}{D}$.
    \item Send the  result `$x$' to Bob who
    applies  a unitary operator $U_x^\dagger$ on B, and thus
    recovers
    the state $|\Psi\rangle$ on subsystem B.
 \end{itemize}

 Now we assume that the shared resource is
$|\varphi^{'}\rangle_{AB}$ given by Eq. \eqref {Ts(i)}. Then we
have
 \begin{align}
 \begin{split}
|\Psi\rangle_a|\varphi^{'}\rangle_{AB}&=(I_a\otimes U_A\otimes
I_B)|\Psi\rangle_a|\varphi\rangle_{AB}\\
&=\frac{1}{\sqrt{D}}\sum_{x=1}^D(I_a\otimes
U_A)|\Phi_x\rangle_{aA}U_x|\Psi\rangle_B.
\end{split}
 \end{align}
The above process shows that if Alice can fulfill her task by
sharing state $|\varphi\rangle_{AB}$, then she can also do that by
sharing state $|\varphi^{'}\rangle_{AB}$, with a projective
measurement on her system `aA' in a basis that includes
$\{(I_a\otimes U_A)|\Phi_x\rangle_{aA}\}$.

Similarly, Alice and Bob can also share the state
$|\varphi^{''}\rangle_{AB}$ given by Eq. \eqref{Ts(ii)}. Then we
have
\begin{align}
\begin{split}
|\Psi\rangle_a|\varphi^{''}\rangle_{AB}&=(I_a\otimes I_A\otimes
V_B)|\Psi\rangle_a|\varphi\rangle_{AB}\\
&=\frac{1}{\sqrt{D}}\sum_{x=1}^D|\Phi_x\rangle_{aA}(V_BU_x)|\Psi\rangle_B.
\end{split}
 \end{align}
This shows that by sharing state $|\varphi^{''}\rangle_{AB}$,
Alice and Bob can fulfill the same task as by sharing state
$|\varphi\rangle_{AB}$, with the only change that Bob's operator
set turns to $\{U_x^\dagger V_B^\dagger\}$.

\subsection{ \label{Protocols:3}Application I: perfect teleportation of qubit states  and superdense coding  of two classical bits via  states of W-class}
Now there are two tasks described below. (i) Can Alice teleport a
state of qubit to Bob by sharing a W-state between them? (ii) Can
Alice send Bob two classical bits of information by sending only
one qubit, with the help of a shared W-state?  In
Ref.~\cite{Joo03}, the authors showed that by  sharing state
$|\widetilde{W}\rangle_{123}$ between Alice and Bob, task (i) can
be fulfilled in a probabilistic manner, but not in a perfect
fashion. Also it is readily seen that the state can not be used
for task (ii) in a perfect fashion. These results seem to imply
that W-states are not suitable for perfect teleportation
 and superdense coding. However, Ref.~\cite{Pan06} found that a class of
 states within W-class can be used to do that. Those states are given
 in the form
 \begin{align}
 \begin{split}
|W_n\rangle_{123}
 =\frac{1}{\sqrt{2+2n}}(&|100\rangle_{123}+\sqrt{n}e^{i\gamma}|010\rangle_{123}\\
 &+\sqrt{n+1}e^{i\delta}|001\rangle_{123}).
 \label{Wn123}
 \end{split}
 \end{align}
A prototype state in this class is  $|W\rangle_{123}$ given by
Eq.~\eqref{3}. In the conclusions of \cite{Pan06}, the authors
proposed some problems worthy of further consideration, one of
which is whether there are other classes of W-states useful for
perfect teleportation and superdense coding. Now with those
transformations we stated previously, indeed we can discover a
class of W-states useful for perfect teleportation and superdense
coding which includes these states given by Eq.~\eqref{Wn123}. As
a result, it will explain why state $|W_n\rangle_{123}$ can be
used for tasks (i) and (ii).

As we know state $|GHZ\rangle_{123}$ is suitable for tasks (i) and
(ii). Here we have a brief review of these protocols for that.
Alice and Bob share state $|GHZ\rangle_{123}$, of which Bob has
particle `3', and Alice has particles `1' and `2'. Also Alice has
particle `a' with state
$|\Psi\rangle=\alpha|0\rangle+\beta|1\rangle$ that will be
teleported from Alice to Bob. Then there is
\begin{align}
\begin{split}
&|\Psi\rangle_a|GHZ\rangle_{123}\\
&=\frac{1}{2}[|\psi^+_1\rangle_{a12}\otimes|\Psi\rangle_3+
|\psi^-_1\rangle_{a12}\otimes\sigma_3|\Psi\rangle_3\\
&\hskip 4mm +|\psi^+_2\rangle_{a12}\otimes\sigma_1|\Psi\rangle_3+
|\psi^-_2\rangle_{a12}\otimes(i\sigma_2)|\Psi\rangle_3],
\end{split}
\end{align}
where $\{|\psi^\pm_1\rangle,|\psi^\pm_2\rangle\}$ are given by
Eq.~\eqref{eight-state} in Sec.~\ref{Protocols:4}. Alice now makes
a projective measurement on particles `a12' in a basis including
$\{|\psi^\pm_1\rangle,|\psi^\pm_2\rangle\}$, and then sends the
measurement results to Bob who can recover state $|\Psi\rangle$ at
particle `3' by applying appropriate operations. Therefore task
(i) is fulfilled.   For task (ii), similarly, $|GHZ\rangle_{123}$
is shared between Alice and Bob, and we let Alice have particle
`3' and let Bob have the left particles. Then Alice first applies
$\{I, \sigma_1, -i\sigma_2, \sigma_3\}$ on her qubit and then
sends her qubit to Bob. As a result, the possible states of the
three qubits holden by Bob will form an orthogonal set with four
states that can be perfectly distinguished by Bob. Therefore Bob
can get two classical bits from Alice.

 Now with the above protocols and  the transformations stated in Sec.~\ref{Protocols:1} and
\ref{Protocols:2}, we can obtain a subclass of W-class  useful for
perfect teleportation and superdense coding. We give our result as
follows.

\textit {Theorem 1.}
  A state of W-class is suitable for perfect teleportation and
superdense coding if, and only if it can be converted from state
$|GHZ\rangle_{123}$ by such a unitary operation that is the tensor
product of a two-qubit unitary operation and a one-qubit unitary
operation.

\begin{figure}
 \setlength{\unitlength}{1.7mm}
 \vskip 2mm
 \begin{picture}(45,35)

\put(0,18){ \put(0.3,17.5){Alice$\Bigg\{$}
 \put(1,10.5){Bob}
\put(7,21){a}
 \put(7,18.5){1}
 \put(7,16){2}
 \put(7,11.5){3}

\put(6,20.5){\line(1,0){12}} \put(6,18){\line(1,0){12}}
\put(6,15.5){\line(1,0){12}} \put(5,11){\line(1,0){23}}

\put(18,15){\framebox(6,6){\qbezier(1,4)(4,4)(5,1)\put(2,1){\vector(1,1){3}}}}

\put(24,18){\line(1,0){6}}\put(26,18.5){$P_x$}
\put(30,18){\vector(0,-1){5}}\put(30.3,15){$x$}

\put(28,9){\framebox(4,4){$U^x_3$}}

 \put(32,11){\line(1,0){6}}

\put(7.2,9){\vector(0,1){2}}\put(5,7){$|\Psi\rangle_a|GHZ\rangle_{123}$}
\put(36,9){\vector(0,1){2}} \put(35,7){$|\Psi\rangle$}
 }

\put(40,30){$\Longrightarrow$}
 \put(0,0){  \put(0.3,17.5){Alice$\Bigg\{$}
 \put(1,10.5){Bob}
\put(7,21){a}
 \put(7,18.5){1}
 \put(7,16){2}
 \put(7,11.5){3}

\put(6,20.5){\line(1,0){12}} \put(6,18){\line(1,0){12}}
\put(6,15.5){\line(1,0){12}} \put(5,11){\line(1,0){23}}

\put(6,20.5){\line(1,0){12}} \put(6,18){\line(1,0){12}}
\put(6,15.5){\line(1,0){12}} \put(5,11){\line(1,0){23}}

\put(18,15){\framebox(6,6){\qbezier(1,4)(4,4)(5,1)\put(2,1){\vector(1,1){3}}}}

\put(24,18){\line(1,0){6}}\put(26,18.5){$P_x^{'}$}
\put(30,18){\vector(0,-1){5}}\put(30.3,15){$x$}

\put(27,9){\framebox(6,4){$U^x_3V_3^\dagger$}}

 \put(33,11){\line(1,0){5}}

\put(7.2,9){\vector(0,1){2}}\put(5,7){$|\Psi\rangle_a|W^{'}\rangle_{123}$}
\put(36,9){\vector(0,1){2}} \put(35,7){$|\Psi\rangle$}
 }
 \end{picture}
 \caption{\label{Fig 1} Teleportation of state $|\Psi\rangle$: when the shared resource is changed from $|GHZ\rangle_{123}$ to
  $|W^{'}\rangle_{123}=(V_{12}\otimes V_3)|GHZ\rangle_{123}$, then the measurement projectors made by Alice are changed from  $\{P_x\}$
  to $\{P_x^{'}:(I_a\otimes V_{12})P_x(I_a\otimes V_{12}^\dagger)\}$, and the operators applied by Bob are changed from $\{U_3^x\}$ to $\{U^x_3V_3^\dagger\}$. }
\end{figure}
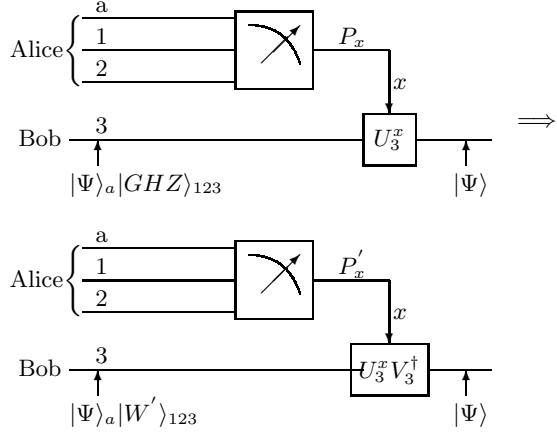

 \textit{Proof.} The sufficiency follows from  the protocols by using $|GHZ\rangle_{123}$ and the transformations stated in Sec.~\ref{Protocols:1} and
\ref{Protocols:2}. We provide  Fig.~\ref{Fig 1} and Fig.~\ref{Fig
2} to show that.

 Next we verify the necessity. From our knowledge about
 teleportation and superdense coding, it follows that a state $|\varphi\rangle_{123}$ of triqubit that can be
 used for tasks (i) and (ii) can necessarily be partitioned into
 two subsystems, say A and B, which share one ebit of
 entanglement. That is, $|\varphi\rangle_{AB}$ can be regarded as a maximally
 entangled state of bisystem AB. In addition, we know that all the maximally
 entangled states of  bisystem AB can be converted from each other
 by local unitary operations (i.e., in the form $U_A\otimes U_B$).
Therefore, the necessity of the theorem follows. This completes
the proof.

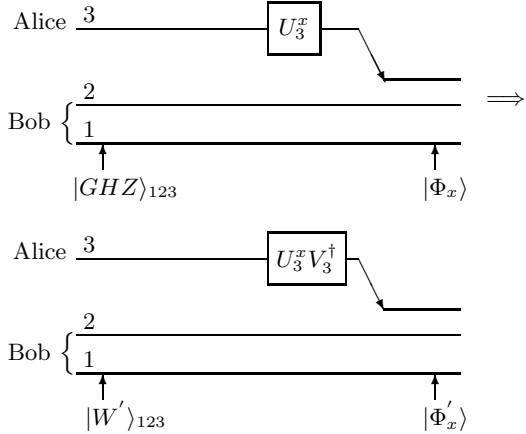
\begin{figure}
 \setlength{\unitlength}{1.7mm}
 \vskip 2mm
 \begin{picture}(40,35)
 \put(0,18){

 \put(3,18){Alice }
 \put(8.5,18.5){3}
 \put(8.5,12.5){2}
\put(8.5,9.5){1} \put(2.6,10.1){Bob $\Big\{$}

 \put(8,18){\line(1,0){15}} \put(23,16){\framebox(4,4){$U_3^x$}}
 \put(27,18){\line(1,0){3}}
  \put(30,18){\vector(1,-2){2.1}}
   \put(32,14){\line(1,0){6}}

 \put(8,12){\line(1,0){30}}
 \put(8,9){\line(1,0){30}}

\put(10,7){\vector(0,1){2}} \put(7,5){ $|GHZ\rangle_{123}$}
\put(36,7){\vector(0,1){2}} \put(35,5){$|\Phi_x\rangle$} }

\put(40,30){$\Longrightarrow$}

 \put(0,0){
  \put(3,18){Alice }
 \put(8.5,18.5){3}
 \put(8.5,12.5){2}
\put(8.5,9.5){1} \put(2.6,10.1){Bob $\Big\{$}

\put(8,18){\line(1,0){15}}
\put(23,16){\framebox(6,4){$U_3^xV_3^\dagger$}}
 \put(29,18){\line(1,0){1}}
  \put(30,18){\vector(1,-2){2}}
   \put(32,14){\line(1,0){6}}

\put(8,12){\line(1,0){30}} \put(8,9){\line(1,0){30}}

\put(10,7){\vector(0,1){2}} \put(8,5){ $|W^{'}\rangle_{123}$}
\put(36,7){\vector(0,1){2}} \put(35,5){$|\Phi_x^{'}\rangle$} }
  \end{picture}
  \caption{ \label{Fig 2} Superdense coding of two classical bits: when the shared resource is changed from $|GHZ\rangle_{123}$ to
 $|W^{'}\rangle_{123}=(V_{12}\otimes V_3)|GHZ\rangle_{123}$,  then the
 operators applied by Alice are changed from $\{U^x_3\}$ to
 $\{U^x_3V_3^\dagger\}$, and the  states resulted  at Bob' end
 are changed from $\{|\Phi_x\rangle\}$ to
$\{|\Phi^{'}_x\rangle:(V_{12}\otimes I_3)|\Phi_x\rangle\}$. }
\end{figure}

Now we  have a brief test  that the  states given by
Eq.~\eqref{Wn123} used in \cite{Pan06} for perfect teleportation
and superdense coding are contained  in the subclass  stated by
us. Firstly we can rewrite Eq.~\eqref{Wn123} in the following
\begin{align}
|W_n\rangle_{123}=\frac{1}{\sqrt{2}}(|\phi\rangle_{12}|0\rangle_3+e^{i\delta}|00\rangle_{12}|1\rangle_{3}),
\end{align}
where
$|\phi\rangle=\frac{1}{\sqrt{1+n}}(|10\rangle+\sqrt{n}e^{i\gamma}|01\rangle)$.
Then $|W_n\rangle_{123}$ can be converted from $|GHZ\rangle_{123}$
by
\begin{align}
|W_n\rangle_{123}=(V_{12}\otimes I_3)|GHZ\rangle_{123}
\end{align}
where $V_{12}$ is a unitary operator acting on particles `1' and
'2' given as
\begin{align}
V_{12}=|\phi\rangle\langle00|+|11\rangle\langle01|+|\phi^\perp\rangle\langle10|+e^{i\delta}|00\rangle\langle11|,
\end{align}
where$|\phi^\perp\rangle=\frac{1}{\sqrt{1+n}}(\sqrt{n}e^{-i\gamma}|10\rangle-|01\rangle)$.
Now we have shown that $|W_n\rangle_{123}$ satisfies the condition
given in Theorem 1. Thus it is natural to use it for teleportation
and superdense coding as did in \cite{Pan06}. In addition,
 as a special state in Eq.~\eqref{Wn123}, $|W\rangle_{123}$ is
of course  suitable for that.

\subsection{\label{Protocols:4}Application II: use state $|W\rangle_{123}$ to transmit three classical bits by sending two qubits}
In Ref.~\cite{Pan06} the authors thought that  it
 may be not possible to use state $|W\rangle_{123}$ to transmit three
classical bits by sending two qubits. Then they conjectured that
there may exist some other states of W-class that can be used for
such a task. In fact, state $|W\rangle_{123}$ can be used to do
that as we will show below. Furthermore, there are indeed many
W-states can be used for that.

 Firstly
we know that state $|GHZ\rangle_{123}$ can be shared by Alice and
Bob, such that Alice can send Bob three classical bits of
information by sending two qubits \cite{Cer01, Pan06}. In this
case, Alice possesses the first two qubits, and the unitary
operators applied by Alice can be in the set
\begin{align}
\begin{split}
{\cal U}_{12}=\{&I\otimes I, \sigma_1\otimes I,(-i\sigma_2)\otimes
I,
\sigma_3\otimes I, \\
&I\otimes \sigma_1, I\otimes (-i\sigma_2), \sigma_1 \otimes
\sigma_1, \sigma_1 \otimes (-i\sigma_2)\}.
\end{split}
\end{align}
After Alice's operation on her qubits and sending that to Bob, the
state of the three qubits will be in one of the following states
\begin{align}
\begin{split}
&|\psi_1^\pm\rangle=\frac{1}{\sqrt{2}}(|000\rangle\pm|111\rangle),\hskip
2mm
|\psi_2^\pm\rangle=\frac{1}{\sqrt{2}}(|100\rangle\pm|011\rangle),\\
&|\psi_3^\pm\rangle=\frac{1}{\sqrt{2}}(|010\rangle\pm|101\rangle),\hskip
2mm
|\psi_4^\pm\rangle=\frac{1}{\sqrt{2}}(|110\rangle\pm|001\rangle).
\end{split} \label{eight-state}
\end{align}
Since the above eight states are mutually orthogonal, Bob can make
a projective measurement to perfectly distinguish them, such that
he can get three classical bits of information from Alice.

 Now from the feasible protocol above, we can obtain that $|W\rangle_{123}$
 is also suitable for transmitting  three classical bits  by sending two qubits. First of all, one should notice that
 \begin{align}
(U_{12}\otimes I_3)|GHZ\rangle_{123}=|W\rangle_{123} \label{GHZ-W}
 \end{align}
where \begin{align}
U_{12}=|\varphi^+\rangle\langle00|+|11\rangle\langle01|+|\varphi^-\rangle\langle10|+|00\rangle\langle11|,
\label{U12}
\end{align}
and
$|\varphi^\pm\rangle=\frac{1}{\sqrt{2}}(|10\rangle\pm|01\rangle)$.
$U_{12}$ is a joint unitary operator acting on the first two
qubits, and can also be  given by   matrix form in the basis
$\{|00\rangle,|01\rangle,|10\rangle,|11\rangle\}$:
\begin{align}
\left(%
\begin{array}{cccc}
  0 & 0 & 0 & 1 \\
  \frac{1}{\sqrt{2}} & 0 & -\frac{1}{\sqrt{2}}& 0 \\
   \frac{1}{\sqrt{2}} & 0 & \frac{1}{\sqrt{2}} & 0  \\
  0 & 1 & 0 & 0 \\
\end{array}%
\right).
\end{align}

\begin{figure}
 \setlength{\unitlength}{1.7mm}
 \vskip 2mm
 \begin{picture}(40,35)
 \put(0,18){

 \put(2,15){Alice$\Bigg\{$}
 \put(2,9){Bob}
 \put(9,18.5){1}
 \put(9,13.5){2}
 \put(9,9.5){3}

 \put(8,18){\line(1,0){15}}
 \put(8,13){\line(1,0){15}}
 \put(7,9){\line(1,0){30.5}}

\put(23,16){\framebox(4,4){$U_1^x$}}
\put(23,11){\framebox(4,4){$U_2^x$}} \put(27,18){\line(1,0){3}}
 \put(27,13){\line(1,0){3}}
  \put(30,18){\vector(1,-2){3}}
 \put(30,13){\vector(1,-2){1.5}}
  \put(33,12){\line(1,0){4.5}}
\put(31.5,10.2){\line(1,0){6}}

\put(9.3,7){\vector(0,1){2}} \put(6,5){$|GHZ\rangle_{123}$}
\put(36,7){\vector(0,1){2}} \put(35,5){$|\Phi_x\rangle$} }

\put(40,30){$\Longrightarrow$}

 \put(0,0){
 \put(2,15){Alice$\Bigg\{$}
 \put(2,9){Bob}
 \put(9,18.5){1}
 \put(9,13.5){2}
 \put(9,9.5){3}

 \put(8,18){\line(1,0){7}}
 \put(8,13){\line(1,0){7}}
 \put(7,9){\line(1,0){30.5}}

\put(9,18){\line(1,0){6}} \put(8,13){\line(1,0){7}}

\put(15,11){\framebox(4,9){$U_{12}^\dagger$}}

\put(19,18){\line(1,0){4}} \put(19,13){\line(1,0){4}}

 \put(23,16){\framebox(4,4){$U_1^x$}}
\put(23,11){\framebox(4,4){$U_2^x$}}

 \put(27,18){\line(1,0){3}}
 \put(27,13){\line(1,0){3}}
  \put(30,18){\vector(1,-2){3}}
 \put(30,13){\vector(1,-2){1.5}}
  \put(33,12){\line(1,0){4.5}}
\put(31.5,10.2){\line(1,0){6}}

\put(9.5,7){\vector(0,1){2}} \put(7,5){$|W\rangle_{123}$}
\put(36,7){\vector(0,1){2}} \put(35,5){$|\Phi_x\rangle$} }
  \end{picture}
\caption   { \label{Fig 3}Change the channel from
$|GHZ\rangle_{123}$ to
 $|W\rangle_{123}$ for superdense coding of three classical bits.}
\end{figure}
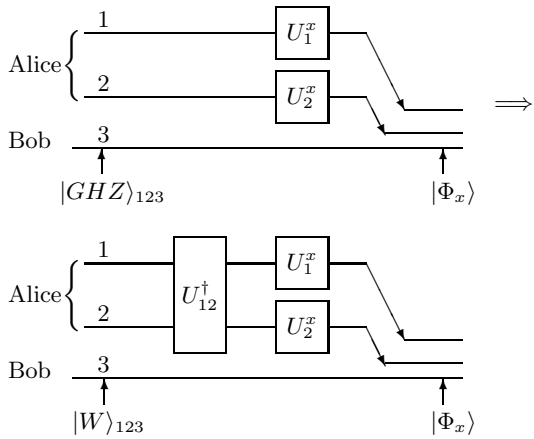

 Now we let Alice and Bob share  state $|W\rangle_{123}$, of
which the first two qubits belong to Alice and the last qubit
belongs to Bob. From Eqs. \eqref{Ts(i)},  \eqref{Tsr(i)} and
\eqref{GHZ-W},  we soon get that if Alice chooses operators
$U_x\in {\cal U}_{12}U_{12}^\dagger$ on her two qubits and then
sends that to Bob,  the eight orthogonal states given by Eq.
\eqref{eight-state} will appear at Bob's end. Therefore, Bob can
make a projective measurement to perfectly distinguish which
operator has been applied by Alice, and then gets three classical
bits of information from Alice. The process is shown in
Fig.~\ref{Fig 3}.

 Note that every operator in the set ${\cal U}_{12}$
can be chosen as  tensor product of two local operators.
Therefore, state $|GHZ\rangle_{123}$ can  be suitable for the case
of two senders and one receiver, where two senders, spatially
separated, make only local operations on their own qubits. This
case was generalized to the case of multi-sender vs one receiver
in \cite{Bos97,Long,Bru05}. With respect to state
$|W\rangle_{123}$, although it can be used to send three classical
bits, it may  be not suitable for the case of two senders and one
receiver, because of the non-locality of $U_x$ that derives from
the non-locality of $U_{12}$. Indeed, it was shown in \cite{Dur00}
that the states in W-class can not be converted from the states in
GHZ-class by local operations. It is also worth pointing out that,
here we just propose a scheme for sending three classical bits
where
 non-local operators are allowed by the sender, but we do not prove the
 impossibility of local operators.

 Note that all the states of W-class that satisfy the condition given
 by Theorem 1 in Sec.~\ref{Protocols:3} are suitable for the task stated in this subsection.

\section{ \label{Characterization}Characterize  W-class of multi-particle systems}

\subsection{ \label{Characterization:1}Generalize states $|W\rangle_{123}$ and $|\widetilde{W}\rangle_{123}$ to multi-qubit systems}
Considering the special properties of states $|W\rangle_{123}$ and
$|\widetilde{W}\rangle_{123}$, it is interesting to  generalize
 them to multi-qubit systems. In fact, it is not difficult to do that. As
 shown in \cite{Dur00}, state $|\widetilde{W}\rangle_{123}$ can be
 easily generalized to multi-qubit  systems in the form ($N\geq 2$):
 \begin{align}
|\widetilde{W}^N\rangle &=\frac{1}{\sqrt{N}}(|\overbrace{10\dots
0}^N\rangle+|01\dots 0\rangle+\dots+|0\dots 01\rangle)\label{WN1}.
 \end{align}
Clearly the state is symmetric in the sense that permutation of
particles does not change state.
  Therefore, if we partition it
into two subsystems in such a way: $(1\dots i-1,i+1\dots N)|i$
(simply denoted by $(\neq i)|i$), then we get the Schmidt
decomposition form:
\begin{align}
|\widetilde{W}^N\rangle=\sqrt{\frac{N-1}{N}}|\widetilde{W}^{N-1}\rangle|0\rangle_i+\sqrt{\frac{1}{N}}|0\dots
 00\rangle|1\rangle_i.
 \end{align}
 Then two density operators associated with the two subsystems are
\begin{align}
\begin{split}
&\rho_i=\frac{1}{N}|1\rangle\langle1|+\frac{N-1}{N}|0\rangle\langle0|,\\
&\rho_{(\neq i)}=\frac{1}{N}|0\dots 0\rangle\langle0\dots
0|+\frac{N-1}{N}|\widetilde{W}^{N-1}\rangle\langle\widetilde{W}^{N-1}|,
\end{split}
\end{align}
where $i=1,2,\dots,N$, and $\rho_{(\neq i)}$ denotes the density
operator of $|\widetilde{W}^{N}\rangle$ losing the $i$th qubit.
Calculating the entanglement between the two subsystems, we have
\begin{align}
E_{(\neq i)|i}(|\widetilde{W}^{N}\rangle) =-p\log_2 p-(1-p)\log_2
(1-p)\leq 1,
\end{align}
 where we let
$p=\frac{1}{N}$. Now from the properties of binary
 entropy \cite{Nie}, we know that the value decreases while $N$ increases, and the equality can be attained
  when $N=2$. In fact, when $N=2$, the state reduces to the EPR pair
  $\frac{1}{\sqrt{2}}(|01\rangle+|10\rangle)$.

State $|W\rangle_{123}$ can be generalized to multi-qubit systems
in the following form:
 \begin{align}
 \begin{split}
|W^N\rangle =&\frac{1}{\sqrt{2(N-1)}}\Big(|\overbrace{10\dots 0}^N
\rangle+\dots+\\
&|0\dots 10\rangle+\sqrt{N-1}|0\dots 01\rangle\Big).
\end{split}
 \end{align}
This state is not fully symmetric. More exactly, permutation of
the first $N-1$ qubits does not change state. If we partition the
state into two subsystems in the way: $(1\dots i-1,i+1\dots N)|i$
as before,
 then when $i\neq N$ it can be rewritten in  the Schmidt decomposition form:
\begin{align}
\begin{split}
 |W^N\rangle=&\sqrt{\frac{2N-3}{2(N-1)}}|\varphi^{N-1}\rangle|0\rangle_i\\
 &+\frac{1}{\sqrt{2(N-1)}}|0\dots
 00\rangle|1\rangle_i,
 \end{split}
 \end{align}
where
\begin{align}
\begin{split}
|\varphi^{N-1}\rangle=&\frac{1}{\sqrt{2N-3}}\Big(|\overbrace{10\dots
0}^{N-1}\rangle+\dots+|0\dots 10\rangle\\
&+\sqrt{N-1}|0\dots 01\rangle\Big).
 \end{split}
\end{align}
Then the entanglement between the two subsystems resulted is
always less than one ebit, and when $N=2$ it attains one ebit.

 When $i=N$,  the state can be rewritten in the form of
maximally entangled state
\begin{align}
 |W^N\rangle=\frac{1}{\sqrt{2}}|\widetilde{W}^{N-1}\rangle|0\rangle_N+\frac{1}{\sqrt{2}}|0\dots
 00\rangle|1\rangle_N.
 \end{align}
Therefore, if we partition $|W^N\rangle$ into two subsystems in
the way: $(1\dots N-1)|N$, then the entanglement between the two
subsystems is: $ E_{(1\dots N-1)|N}(|W^N\rangle)=1$. Thus the
state can be used as a shared resource for teleportation and
superdense coding, which will be detailed in the following
subsection.
\subsection{\label{Characterization:2} State $|W^N\rangle$  as a
shared resource for teleportation and superdense coding}

Here we see how  state $|W^N\rangle$ is used  for  teleportation
and superdense coding.
\paragraph{Teleportation} Let Alice and Bob share  state
$|W^N\rangle$.  Now Alice possesses the first $N-1$ qubits while
Bob possesses the last qubit. Alice also has a particle `a' in the
unknown state $|\Psi\rangle_a=\alpha|0\rangle_a+\beta|1\rangle_a$.
Then the combined input state can be rewritten as
\begin{widetext}
\begin{align}
\begin{split}
|\Psi\rangle_a|W^N\rangle&=(\alpha|0\rangle_a+\beta|1\rangle_a)\frac{1}{\sqrt{2}}(|\widetilde{W}^{N-1}\rangle|0\rangle+|0\dots
 00\rangle|1\rangle)\\
 &=\frac{1}{\sqrt{2}}\Big[ \alpha|0\rangle_a|\widetilde{W}^{N-1}\rangle|0\rangle+\alpha|0\rangle_a|0\dots
 00\rangle|1\rangle+\beta|1\rangle_a|\widetilde{W}^{N-1}\rangle|0\rangle+\beta|1\rangle_a|0\dots
 00\rangle|1\rangle\Big]\\
 &=\frac{1}{2}\Big[(|\eta^+\rangle+|\eta^-\rangle)\alpha|0\rangle+(|\xi^+\rangle-|\xi^-\rangle)\alpha|1\rangle+(|\xi^+\rangle+|\xi^-\rangle)\beta|0\rangle+(|\eta^+\rangle-|\eta^-\rangle)\beta|1\rangle\Big]\\
 &=\frac{1}{2}\Big[|\eta^+\rangle(\alpha|0\rangle+\beta|1\rangle)+|\eta^-\rangle(\alpha|0\rangle-\beta|1\rangle)+|\xi^+\rangle(\alpha|1\rangle+\beta|0\rangle)+|\xi^-\rangle(\beta|0\rangle-\alpha|1\rangle)\Big],
\end{split}
\end{align}
\end{widetext}
where $\{|\eta^\pm\rangle,|\xi^\pm\rangle\}$ is a set of
orthogonal states given by
\begin{align}
\begin{split}
&|\eta^\pm\rangle=\frac{1}{\sqrt{2}}\Big(|0\rangle_a|\widetilde{W}^{N-1}\rangle\pm
|1\rangle_a|0\dots 00\rangle\Big),\\
&|\xi^\pm\rangle=\frac{1}{\sqrt{2}}\Big(|1\rangle_a|\widetilde{W}^{N-1}\rangle\pm
|0\rangle_a|0\dots 00\rangle\Big).
\end{split}
\end{align}
 Now Alice makes a projective measurement in a basis
that includes the states $\{|\eta^\pm\rangle,|\xi^\pm\rangle\}$ on
the particles possessed by herself. Then she sends the results of
her measurement using two classical bits to Bob who can apply one
of the unitary operators $\{I, \sigma_3, \sigma_1, -i\sigma_2\}$
to convert the state of his particle to that of particle `a'. Now
the teleportation protocol has been completed. This protocol
consumes one ebit of shared entanglement and two bits of classical
communication between Alice and Bob.
\paragraph{Superdense coding} Let Alice and Bob share  state
$|W^N\rangle$. If Bob possesses the first $N-1$ qubits and Alice
possesses the last qubit, then the coding protocol is described
below:
\begin{align}
\begin{split}
&00:\hskip 1mm |W^N\rangle\hskip 1mm \underrightarrow{{I^{\otimes
N-1}\otimes I}}\hskip 1mm
\frac{1}{\sqrt{2}}|\widetilde{W}^{N-1}\rangle|0\rangle+\frac{1}{\sqrt{2}}|0\dots
 00\rangle|1\rangle,\\
&01:\hskip 1mm |W^N\rangle\hskip 1mm \underrightarrow{I^{\otimes
N-1}\otimes \sigma_1}\hskip 1mm
\frac{1}{\sqrt{2}}|\widetilde{W}^{N-1}\rangle|1\rangle+\frac{1}{\sqrt{2}}|0\dots
 00\rangle|0\rangle,\\
&10:\hskip 1mm |W^N\rangle\hskip 1mm \underrightarrow{I^{\otimes
N-1}\otimes -i\sigma_2}\hskip 1mm
\frac{1}{\sqrt{2}}|\widetilde{W}^{N-1}\rangle|1\rangle-\frac{1}{\sqrt{2}}|0\dots
 00\rangle|0\rangle,\\
 &11:\hskip 1mm |W^N\rangle\hskip 1mm \underrightarrow{I^{\otimes
N-1}\otimes \sigma_3}\hskip 1mm
\frac{1}{\sqrt{2}}|\widetilde{W}^{N-1}\rangle|0\rangle-\frac{1}{\sqrt{2}}|0\dots
 00\rangle|1\rangle.
 \end{split}
\end{align}
Alice applies the operators selected from $\{I, \sigma_1,
-i\sigma_2, \sigma_3\}$ on her qubit. One can readily find that
the four states produced are mutually orthogonal. Now Alice can
send her qubit to Bob who makes a projective measurement on the N
qubits in a basis that includes the four states produced. Since
the states produced are orthogonal, Bob can perfectly distinguish
which operator Alice has applied. Therefore, Bob can decode two
classical bits of information from Alice's encoding.

 Note that the two protocols  by using state $|W\rangle_{123}$
stated in \cite{Pan06} are the special case of the above
generalized protocols when $N=3$.

\subsection{ \label{Characterization:3}The optimal way to partition state
$|\widetilde{W}^N\rangle$ into two subsystems} As we can see,
state $|\widetilde{W}^N\rangle$ has a good property---symmetry
that is interesting and useful for some applications. Thus this
state may be worthy of further consideration. In the previous
subsection, we investigated the case of partitioning it into two
subsystems such that one subsystem possesses one qubit and the
other subsystem possesses the residual qubits, and then we found
that the entanglement between the two subsystems becomes close to
0 when $N$ increases, which may be not a good news for our
teleportation and superdense coding. However, if we make light of
this state just because of these results, then that may be the
really bad news. In the following we will find an optimal
partitioning way in the sense  that the entanglement between the
two subsystems resulted attains  its maximum value.

 A general way of partitioning state $|\widetilde{W}^N\rangle$ into two subsystems is to let
 one subsystem have $x$ ($1\leq x<N$ and $x$ is an integer number) qubits and the other subsystem
 the left
 $N-x$ qubits. Then, due to its symmetry property, no matter what  the $x$ qubits are,
 the state can always be rewritten in the form:
 \begin{align}
 \begin{split}
|\widetilde{W}^N\rangle&=\frac{1}{\sqrt{N}}\Big[\Big(|\overbrace{10\dots
0}^x\rangle+\dots +|0\dots 01\rangle\Big)|\overbrace{0\dots
0}^{N-x}\rangle\\
&\hskip 3mm +|\overbrace{0\dots
0}^{x}\rangle\Big(|\overbrace{10\dots 0}^{N-x}\rangle+\dots
+|0\dots
01\rangle\Big)\Big]\\
&=\sqrt{\frac{x}{N}}|\widetilde{W}^x\rangle|\overbrace{0\dots
0}^{N-x}\rangle+\sqrt{\frac{N-x}{N}}|\overbrace{0\dots
0}^x\rangle|\widetilde{W}^{N-x}\rangle
\end{split}
 \end{align}
where states $|\widetilde{W}^x\rangle$ and
$|\widetilde{W}^{N-x}\rangle$ are resulted from $N$ in Eq.
\eqref{WN1}  being $x$ and $N-x$, respectively. Then the
entanglement between the two subsystems resulted, say $A$ and $B$,
is the binary entropy given by
\begin{align}
E_{A|B}(|\widetilde{W}^N\rangle)=-\frac{x}{N}\log_2\frac{x}{N}-(1-\frac{x}{N})\log_2(1-\frac{x}{N}).\label{E(AB)}
\end{align}
Thus from the properties of binary entropy,   we know that the
absolute value $|\frac{x}{N}-\frac{1}{2}|$  is smaller, the value
in Eq. \eqref{E(AB)} is bigger, and the maximum value 1 is
attained when $\frac{x}{N}=\frac{1}{2}$. Therefore, there are two
cases we should consider:
\begin{itemize}
    \item (i) $N$ is even. Then the maximum value 1 can always be attained
    by letting $x=\frac{N}{2}$. That is,  in this case, a balanced partition always leads to
    two subsystems sharing 1 ebit of entanglement.
    \item (ii) $N$ is odd. In the case, we let $x=\lfloor
    \frac{N}{2}\rfloor$, that is, $x$ is the integer part of
    number $ \frac{N}{2}$. Then the value $|\frac{x}{N}-\frac{1}{2}|$
    is the
    smallest among all possible $x$, and thus
    $E_{A|B}(|\widetilde{W}^N\rangle)$ attains its maximum value among all possible
    $x$. Of course, in this case, the entanglement can not attain 1
    ebit
    forever, but when $N$ is large enough, the value can
    arbitrarily close to 1.
\end{itemize}

In summary, for state $|\widetilde{W}^N\rangle$, among all the
ways of partitioning it into two subsystems, the optimal  way is
such one that leads to two balanced (or approximately balanced)
subsystems. Below we provide an example to apply the way.

\subsubsection{An example: state $|\widetilde{W}^4\rangle$ used for teleportation of entangled pairs} As an example, we consider state
$|\widetilde{W}^N\rangle$  for $N=4$. That is the following state
\begin{align}
|\widetilde{W}^4\rangle&=\frac{1}{\sqrt{4}}(|1000\rangle+|0100\rangle+|0010\rangle+|0001\rangle).
\end{align}
  At the first blush, it seems impossible to be comparable with the
  following maximally entangled
    state
\begin{align}
|GHZ^4\rangle=\frac{1}{\sqrt{2}}(|0000\rangle+|1111\rangle).
\end{align}
  However, according to what we have obtained before,  if we
partition it into two subsystems such that every subsystem has two
qubits, then we can rewrite it in this form:
\begin{align}
|\widetilde{W}^4\rangle=\frac{1}{\sqrt{2}}(|\varphi^+\rangle|00\rangle+|00\rangle|\varphi^+\rangle)
\end{align}
 where
$|\varphi^\pm\rangle=\frac{1}{\sqrt{2}}(|10\rangle\pm|01\rangle)$.
In this way, the entanglement between the two subsystems is one
ebit. At the same time, note  that if we partition $|GHZ^4\rangle$
in the same way, then the entanglement between two subsystems
resulted is also one ebit. Therefore, in a certain situation,
$|\widetilde{W}^4\rangle$ can be used to fulfill the same task as
by using $|GHZ^4\rangle$.

Teleportation of entangled pairs via three-particle states was
discussed in \cite{Mar00,Gor00,Gor03}.  Next we see how state
$|\widetilde{W}^4\rangle$ is used as a shared resource for
teleportation of an entangled pair. Detailedly, Bob and Charlie
possess the $3$th and $4$th qubits, respectively, and Alice
possesses the first two qubits. Alice also possesses an entangled
pair of qubits given by
\begin{align}
|\psi\rangle_{ab}=\alpha|00\rangle+\beta|11\rangle, \hskip
2mm\text{or}\hskip 2mm
|\psi^{'}\rangle_{ab}=\alpha|01\rangle+\beta|10\rangle.
\end{align}
Then we have
\begin{align}
\begin{split}
&|\psi\rangle_{ab}|\widetilde{W}^4\rangle\\
&=(\alpha|00\rangle+\beta|11\rangle)\frac{1}{\sqrt{2}}(|\varphi^+\rangle|00\rangle+|00\rangle|\varphi^+\rangle)\\
&=\frac{1}{2}\Big[
|\mu^+\rangle(\alpha|00\rangle+\beta|\varphi^+\rangle)
+|\mu^-\rangle(\alpha|00\rangle-\beta|\varphi^+\rangle)\\
&+|\omega^+\rangle(\alpha|\varphi^+\rangle+\beta|00\rangle)
+|\omega^-\rangle(\alpha|\varphi^+\rangle-\beta|00\rangle) \Big],
\end{split}
\end{align}
and similarly,
\begin{align}
\begin{split}
&|\psi^{'}\rangle_{ab}|\widetilde{W}^4\rangle\\
&=(\alpha|01\rangle+\beta|10\rangle)\frac{1}{\sqrt{2}}(|\varphi^+\rangle|00\rangle+|00\rangle|\varphi^+\rangle)\\
&=\frac{1}{2}\Big[|\pi^+\rangle(\alpha|00\rangle+\beta|\varphi^+\rangle)
+|\pi^-\rangle(\alpha|00\rangle-\beta|\varphi^+\rangle)\\
&+|\varpi^+\rangle(\alpha|\varphi^+\rangle+\beta|00\rangle)
+|\varpi^-\rangle(\alpha|\varphi^+\rangle-\beta|00\rangle)
 \Big],
\end{split}
\end{align}
 where
\begin{align}
\begin{split}
&|\mu^\pm\rangle=\frac{1}{\sqrt{2}}(|00\rangle|\varphi^+\rangle
\pm|11\rangle|00\rangle),\\
&|\omega^\pm\rangle=\frac{1}{\sqrt{2}}(|00\rangle|00\rangle
\pm|11\rangle|\varphi^+\rangle),\\
&|\pi^\pm\rangle=\frac{1}{\sqrt{2}}(|01\rangle|\varphi^+\rangle\pm10\rangle|00\rangle),\\
&|\varpi^\pm\rangle=\frac{1}{\sqrt{2}}(|01\rangle|00\rangle\pm|10\rangle|\varphi^+\rangle).
\end{split}
\end{align}
 The set
$S=\{|\mu^\pm\rangle,|\omega^\pm\rangle,|\pi^\pm\rangle,|\varpi^\pm\rangle\}$
can be extended to an orthogonal basis of the Hilbert space
$H_2^{\otimes4}$. Now Alice can make a projective measurement in a
basis that includes $S$ on the four qubits possessed by her, and
sends the results of measurement using three  cbits to Bob and
Charlie. Then Bob and Charlie can together choose an appropriate
joint unitary operator (non-local operator) on their qubits to
recover the state $|\psi\rangle_{ab}$ or $|\psi^{'}\rangle_{ab}$
at their qubits. For instance, if the result of measurement is
$\mu^+$, then Bob and Charlie make a unitary operation
$U=|00\rangle\langle00|+|\varphi^+\rangle\langle11|+|11\rangle\langle\varphi^+|+|\varphi^-\rangle\langle\varphi^-|$
on their two qubits.

Note that in the above protocol, the two kinds of states
$|\psi\rangle_{ab}$ and $|\psi^{'}\rangle_{ab}$ can be teleported
by using the same protocol with three bits of classical
communication. Here if we just want to teleport one kind of them,
say $|\psi\rangle_{ab}$, then two bits of classical communication
is enough. In addition, since the operations done by Bob and
Charlie are non-local, Bob and Charlie can not be spatially
separated, which may be the limitation of this protocol. On the
other hand, because of this non-locality, Bob and Charlie must
cooperate to recover state $|\psi^{'}\rangle_{ab}$ or
$|\psi\rangle_{ab}$, so this protocol may be considered for {\it
quantum secret sharing} \cite{Hil98}. By the way, state
$|\widetilde{W}^4\rangle$ can also be used  to transmit  three
classical bits by  sending two qubits.
\subsection{ \label{Characterization:4}Generalize state $|W\rangle_{123}$ to multi-particle systems  with higher dimension} In the following, we consider the state of multi-particle
systems with $d$-dimensional particles, such that it generalizes
the state $|W\rangle_{123}$. First we let
\begin{align}
|\xi^N_{(i)}\rangle=\frac{1}{\sqrt{N}}\big(|\overbrace{i0\dots0}^N\rangle+|0i\dots
0\rangle+\dots |0\dots0i\rangle\big)
\end{align}
for $i=1,\dots,d-1$.  Then we let
\begin{align}
|\Omega^N\rangle=\frac{1}{\sqrt{d}}\Big[\sum^{d-1}_{i=1}|\xi^{N-1}_{(i)}\rangle|i-1\rangle+|0\dots0\rangle|d-1\rangle\Big].
\label{N-d}
\end{align}
Now  state $|\Omega^N\rangle$ may be taken as a reasonable
generalization of state $|W\rangle_{123}$. In fact, Eq.
\eqref{N-d} can be regarded as the Schmidt decomposition  for the
state of bisystem AB, where subsystem A consists of the first
$N-1$ particles and subsystem B has the $N$th particle. Thus, the
entanglement between them is $\log_2 d$. Clearly, when $d=2$,
state $|\Omega^N\rangle$ reduces into state $|W^N\rangle$ given by
Eq.~\eqref{WN1}.

Next we see how state $|\Omega^N\rangle$ is used for superdense
coding. Suppose that Alice and Bob share state $|\Omega^N\rangle$
such that Alice possesses subsystem B and Bob possesses subsystem
A.  Alice can now manipulate her subsystem by unitary operations
\begin{align}
 U(m,n)=\sum^{d-1}_{k=0}e^{2\pi ikm/d}|k\rangle\langle
k\oplus n |
 \end{align}
where $\oplus$ denotes addition modulo $d$, and $m,n=0,\dots,
d-1$. After that, Alice sends her subsystem to Bob.  Then similar
to the case in \cite{Long}, $d^2$ mutually orthogonal states will
be produced at the disposal of Bob who can make a projective
measurement on the $N$ particles to distinguish what operation
Alice has applied. Therefore, Bob  gets $2\log_2 d$ classical bits
of information from Alice.

\section{\label{Con}conclusion}
In this work,  we investigated  that the states of W-class are
used for teleportation and superdense coding, and  characterized
W-class in multi-particle systems.
   We  described two transformations of the shared resources for teleportation and superdense
  coding, with which we obtained a
sufficient and necessary condition for a state of W-class being
suitable for perfect teleportation and superdense coding, and
 we found that the state
  $|W\rangle_{123}$
  can be used to transmit three classical bits by sending two qubits, which was considered to be impossible in \cite{Pan06}.
  We generalized the states of
  W-class to multi-qubit systems and multi-particle systems with  higher
  dimension. We proposed two protocols for teleportation
  and superdense coding by using W-states of multi-qubit systems that generalize the protocols by using
  $|W\rangle_{123}$ stated in \cite{Pan06}. We obtained an optimal way to partition   some W-states
  of multi-qubit systems  into two subsystems, such that the entanglement between them achieves
  maximum value.

As we pointed out in Sec.~\ref{Protocols:4}, although state
$|W\rangle_{123}$  can be used to transmit three classical bits by
sending two qubits,  it may be not suitable for the case of two
senders and one receiver. Then in future, one can consider whether
there exist W-states suitable for superdense coding with two
senders and one receiver. If the answer is yes, then one can
further generalize that to the case of multi-sender and one
receiver. Otherwise, one should prove the impossibility.

  This work is
supported in part by the National Natural Science Foundation (Nos.
90303024, 60573006), the Higher School Doctoral Subject Foundation
of Ministry of Education (No. 20050558015), and the Natural
Science Foundation of Guangdong Province (No. 031541) of China.

\end{document}